# The magnetospheric chaos hypothesis:
# a new point of view of the magnetospheric dynamics *

### Historical evolution of magnetoshperic chaos hypothesis the past two decades

G.P. Pavlos

Department of Democritus University of Thrace - Greece

**Abstract**

This study was announced for the first time in Greek language in 1988 in the National Observatory of Athens, in the conference of the 1st Panhellenic Symposium entitled "SOLAR AND SPACE RESEARCH IN GREECE TODAY - Basic Research, Technology and Applications" organized by the National Observatory of Athens, in Penteli (Athens) in 1988. It was also republished in the Conference proceedings of the 2nd Panhellenic Symposium held in Democritus University of Thrace, Xanthi - Greece, 26-29 April 1993. In this study, G.P. Pavlos, initially inspired from the theory of self organization of Prigogine [1] and Nicolis [2], proposes the hypothesis of magnetospheric chaos believing that the classical theory of plasma statistics and plasma instabilities are not able to describe efficiently the holistic character of magnetospheric dynamics. The translation of this study in English language is significant concerning the history of scientific ideas, since the research that took place the following years vindicated the hypothesis of magnetospheric chaos which this study first introduced in 1988.

**Recent evolutions concerning the magnetospheric chaos**

Afterwards, Baker [3] introduces the chaotic modeling of dripping faucet fitting it into magnetospheric dynamics. Moreover, T. Chang introduces powerful theoretical tools and concepts, like renormalization theory and multiscale multifractal process in relation to chaos. Till then there was a series of studies concerning the hypothesis of magnetospheric chaos which supported the concept of a magnetospheric strange attractor (Vassiliadis [4], Pavlos [5] and others).

In parallel, a serious criticism was developed against the the hypothesis of magnetospheric chaos towards two directions. The first important objection concerned the

---

discrimination possibility between signals of colored noise and low dimensional deterministic chaos. Many scientists were discouraged from this negative criticism and drew away from magnetospheric chaos hypothesis, considering that the magnetospheric signals correspond to a form of pseudochaos and to colored noise. In order to answer to this criticism, G.P. Pavlos, in a series of studies established faithfully the magnetospheric chaos developing a preliminary methodology for the discrimination between colored noise and deterministic chaos. In this effort, G.P. Pavlos, was essentially assisted by the Greek scientist A. Tsonis. So, in the period 1990-1999 G.P. Pavlos in collaboration with L. Karakatsanis, M. Athanasiou, D. Kugiumtzis and other, created a solid theoretical framework and algorithms which made possible the discrimination between stochasticity and chaoticity. The results of this methodology showed faithful evidence for the existence of magnetospheric chaos and were published in various scientific journals [5]. These tools and algorithms were also used from G.P. Pavlos and its collaborates (A. Iliopoulos, L. Karakatsanis and others) in order to trace deterministic chaos also in other complex dynamical systems (Solar, Earthquakes, Human brain) apart from magnetospheric dynamics [6-9]. The second serious objection against magnetospheric chaos came from the rapidly growth and acceptance of Self Organized Criticality (SOC) (anafores) by many scientists. The theories of SOC and Chaos were considered initially two contradicting theories. However, G.P. Pavlos and the Group of Thrace in a recent series of studies [10-12] illustrated through experimental time series that in the space plasma as well as in other complex distributed systems, the high dimensional SOC and the low dimensional Chaos, do not contradict each other, oppontely but correspond to two different phases-states of the same system, which the system manifests simultaneously or sequentially in time through a nonequilibrium phase transition process. Moreover, the results were in agreement with the broader statistical theory concerning systems far from equilibrium described by Chang [13], Consolini [14], Sharma [15], Sitnov [16] and other scientists.

Recently, the Group of Thrace attempts to compile a synthesis of chaos, turbulence and SOC in the theoretical framework of Tsallis theory [10, 11]. Furthermore, in a series of studies, which are under preparation, the dynamics of space plasma is placed between a broader theoretical framework of complexity theory, connected with the fractal generalization of dynamics, scale relativity, strange kinetic theory and the fractal form of space –time according to Nottale [17], Castro [18], Coldfain [19], El Naschie [20], Cresson [21], Zelenyi [22], Milovanov [23] and other scientists.

Recently, Pavlos [10], following new theoretical evolutions of non-equilibrium complexity theory introduced the concepts of fractal dissipation and fractal acceleration for the physical understanding of the magnetospheric dynamics.

Finally, it is worth to mention that the recent progress, for about two decades after this study, signified that the magnetospheric system and its chaotic dynamics comprised a basic example of the application of chaos and complexity to other regions of space plasma, as well as to other complex systems such as seismogenesis or human brain, etc. At the same time, space plasma dynamics can play a basic role in the understanding of new theoretical ideas under progress, such as fractal generalization of Dynamics, strange kinetics and anomalous diffusion which in general are the manifestation of the self organization process and the emergence of order at far from equilibrium plasma states. In far from equilibrium states, the macroscopic complexity seems to be directly related to microscopic complexity in a way that is still unclear but justifying the theoretical propositions stated by various scientists such as Prigogine, Nicolis, El Naschie, Nottale, Castro and others. This synoptic presentation shows the historical value of these first ideas for magnetospheric chaos which were presented from G.P. Pavlos in 1988, as well as the importance of the significant and novel concept of self organisation process in space plasmas.

In the following, we present, in English translation, the hypothesis of magnetospheric chaos, as it was stated in 1988 (at the Proceedings of the 1st Panhellenic Symposium on SOLAR AND SPACE RESEARCH IN GREECE TODAY, Athens 1988). These ideas preserve until today their scientific interest and importance, since till today there is no unique theoretical description of the space plasmas dynamics.

**Introduction**

*One of the most important goals in space physics is to understand how the basic elements work in the solar wind-magnetosphere-ionosphere interaction. The main manifestation of this interaction are magnetospheric substorms, a collective and global response of the magnetosphere and ionosphere to a set of conditions in the solar wind. Characteristically we mention some of the most typical events during a substorm expansive phase, such as: sudden brightening of equatoward auroral, intensification of field aligned currents and of the westward-eastward auroral electrojet currents, plasma sheet thinning, earthward injection of energetic electrons and ions, energetic bursts in magnetotail, high–speed bursty earthward-tailward plasma flow (500-1000 km/sec) in the central plasma sheet, intensification of MHD waves in different region of the magnetosphere and plasma turbulence.*

*In this work we support the novel concept that the dynamics of the magnetopsheric system can be explained by the development of a magnetospheric chaotic attractor. So, the motion of the magnetospheric state on a chaotic (strange) attractor can be considered as a holistic explanatory paradigm of the global magnetospheric dynamics.*

*This concept is not unreasonable since modern physics in its three stages: relativity, quantum mechanics and non-linear (far from equilibrium) thermodynamics, has revealed that any physical theory is mainly a model and a hypothesis about physical reality concerning its appearance but not its essence. From this point of view the scientific models have to be in coherence*

*with observational data but this cannot exclude the acceptance of new complementary theoretical assumptions.*

*Thus, from the point of view of chaos and nonlinear thermodynamics, the magnetospheric system is an open system, as it exchanges mass and energy with the ionosphere and the solar wind, while it remains far from thermodynamic equilibrium. For this reason the magnetosphere belongs to the class of dissipative chaotic systems. An important consequence of chaos theory for these systems is the possibility of existence of strange attractors in the dynamical phase space. When a system is in the state of a strange attractor, a small perturbation of the system can be amplified and can lead the system to instability and to divergence from its initial state. In disagreement to the original Landau's turbulence theory, chaos theory shows that even for low degrees of freedom the existence of turbulent (chaotic) attractors is possible as well. Our failure until now for developing a sufficient and of local-character description of magnetospheric dynamics, though not extinguishing the hope for future achievements, supports our suggestion that chaos theory may constitute a powerful tool for a global comprehension of magnetospheric dynamics. According to this concept, the central question to be answered through chaos theory is how much the transition of the magnetospheric system from quiet state to the growth phase and subsequently to the explosive phase of substorms corresponds to a transition from a state of simple attractor as a limit point to a chaotic or strange attractor.*

### *Chaos Theory and Magnetospheric Dynamics*

*The hypothesis of the holistic behavior of the magnetospheric system can be related to recent results concerning Chaos theory concerning conservative or dissipative systems [24]. Chaos theory has already been used in the magnetospheric physics, regarding the chaotic behavior of the electron's population according with the tear mode instability [25]. However, in this study we will study the relationship between the magnetospheric dynamics and chaotic physics using another perspective. The study of complex physical systems for example physicochemical, has shown, already, the significance of the holistic character which is manifested when the system is driven far from thermodynamical equilibrium [26].*

*The state of a dynamical system is described by the equations:*

$$\frac{\partial \mathbf{X}_i(t)}{\partial t} = F(X_1, X_2, \ldots, X_n, \lambda), \ i = 1, \ldots, n, \ k = 1, 2, \ldots, m \qquad (1.1)$$

*Where $\mathbf{X}_i$ are the macroscopic or microscopic quantities of the system and $\lambda_k$ are the m control parameters of the system. These equations are known as "flow" equations in the phase space $\Gamma \equiv \{\mathbf{X}_i\}$ and they describe integrable Hamiltonian systems or dissipative open systems which exchange matter and energy with their environment. Far from equilibrium and for critical values of the parameters $\lambda_k$ bifurcations of the solutions of the equations 1.1 take place. In the bifurcation points, microscopic fluctuations (instability development) are amplified with a stochastic way leading the system into new physical states.*

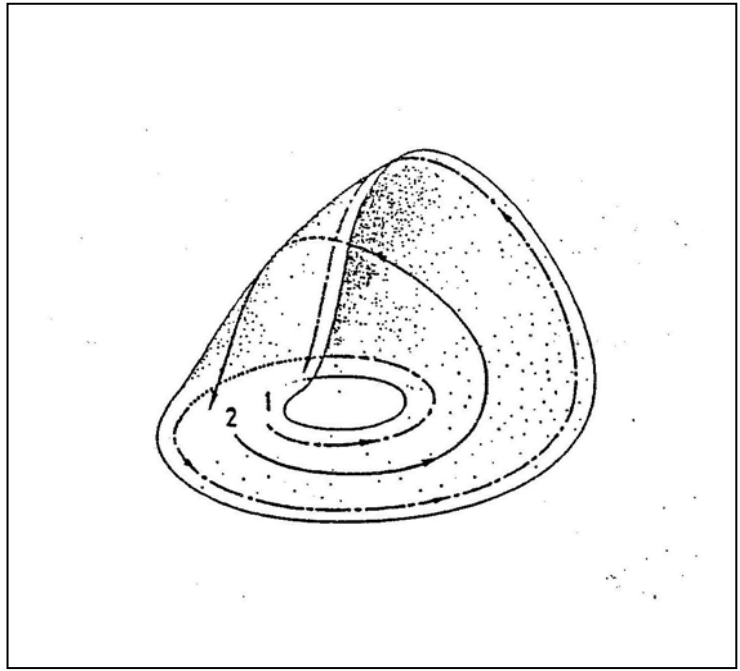

*Figure I: Typical evolution of nearby trajectories in Rössler's attractor.*

*According to the previous paragraph, the magnetospheric system is open and exchanges matter and energy with the ionosphere and the solar wind, staying far from equilibrium. This characteristic places the magnetosphere among the dissipative chaotic systems, for which, a significant consequence of Chaos theory, the existence of low dimensional attractors in the phase space $\Gamma \equiv \{\mathbf{X}_i\}$ is possible. Depending on the system and on the parameter's values (which usually determine the distance from the thermodynamic equilibrium), the attractors are categorized as point attractors, limit cycle or torus attractors (simple attractors), and chaotic or strange attractors with intense stochastic character. Figure 1 shows the shape of a chaotic attractor known as Rossler attractor. The main characteristic of the attractors is that they attract the dynamics defining the solution of the system in their region. Especially, in the chaotic or strange attractor, together with the chaotic character due to instability, there is sensitivity to initial conditions. This causes the attraction of the system dynamics from the basin of the attractor to the attractor region and characterize the asymptotic dynamics on the attractor. In the case of the chaotic attractor, a small perturbation (noise) can be amplified, leading the system into instability and removal from its initial state. In the cases of simple attractors, the autocorrelation of the variables $C(\tau)$*

$$C(\tau) = \frac{1}{t_2 - t_1} \int_{t_1}^{t_2} X(t) X(t+\tau) d\tau \qquad (1.2)$$

*remains constant (and different from zero) indicating, according to Wiener-Khintchine theorem, periodic or semiperiodic Fourier power spectrum with discrete frequencies $\omega_i$.*

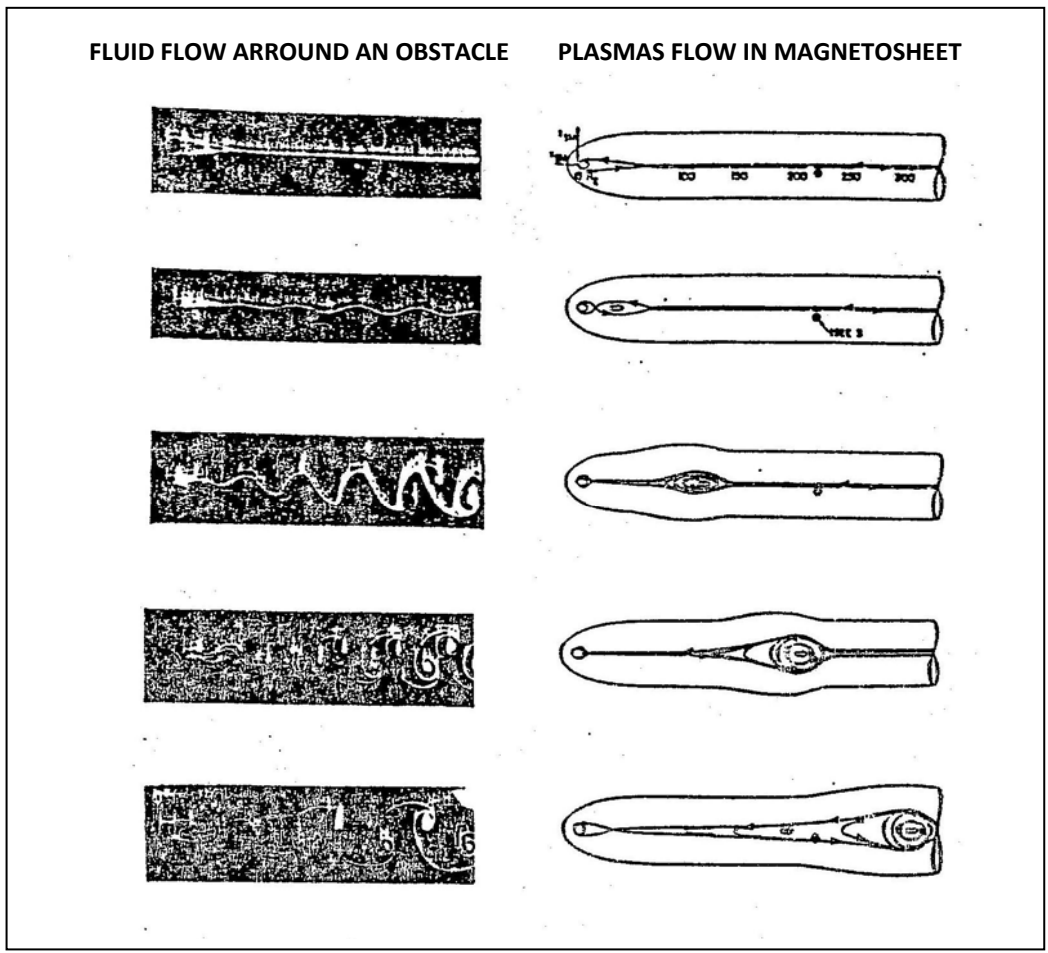

*Figure II: Five snapshots of the evolution of the hydrodynamic flow behind a block for increasing flow velocity (a) and five sequential phases of Earth's magnetotail during magnetic substorms, according to Hones model (b). The similarity between the evolution from laminar to turbulent flow and the evolution from the "quiet" magnetotail to the turbulent flow of the magnetospheric plasma, is obvious.*

*On the opposite, in the case of strange attractors, the autocorrelation function is nullified as $r \to \infty$, while the Fourier transformation of the power spectrum indicates a state of turbulence with a continuous frequency spectrum. In contrast to the Landau turbulence theory [27], in chaos theory can be proven that also in the case of systems with few degrees of freedom it is possible the development of chaotic state and thus the existence of chaotic attractors. Indeed, after two or three bifurcations of the solutions of equations 1.1 chaotic attractors appear.*

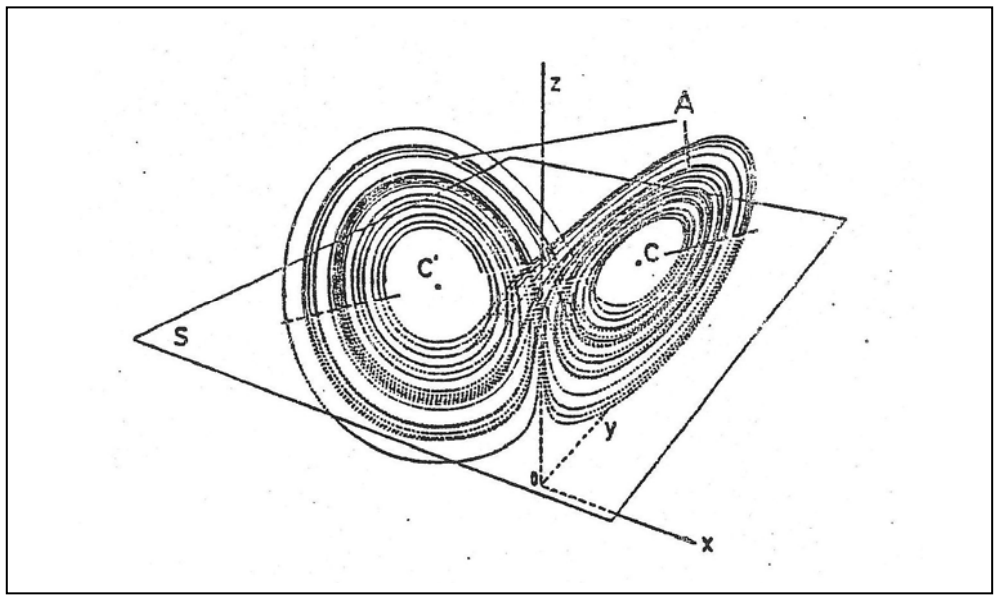

*Figure III: Characteristic shape of the strange (chaotic) Lorenz attractor for parameter values 10.28 and 8.3.*

*Apart from the attractors (simple or chaotic) an additional characteristic of chaos theory is the possibility of the transformation of the partial differential equations which describe a continuous physical system (e.g. a fluid) into equations of the form of 1.1, that can be easily be studied as far as the existence of attractors is concerned (R. Sagdeev et al., 1988). Characteristic examples of this transformation are the Rayleigh-Benard convection and Lorenz model. In these cases, the form of the hydrodynamic equations which were studied, aren't very different from corresponding equations concerning the magnetohydrodynamics of the magnetospheric physics. This relation is shown in Figure 2, where significant analogies can be seen concerning the hydrodynamic flow around a block and the magnetospheric plasma flow in Earth's magnetospheric tail, as the external coupling is amplified and therefore the turbulent character. In the Lorenz model the equations of the hydrodynamic flow are transformed into:*

$$\frac{dX}{dt} = aY - aX$$

$$\frac{dY}{dt} = -XZ + bX - Y \qquad (1.3)$$

$$\frac{dZ}{dt} = XY - cZ$$

*according to which the physical system is treated holistically, without taking into consideration the intrinsic processes of its components.*

*For specific values of the equation parameters the existence of a chaotic attractor is possible, as shown in fig. 3, with the two characteristics of the chaotic attractors being obvious: the sensitivity to initial conditions and the instability of the nearby trajectories. However, the analogy of the simple hydrodynamic models with the magnetospheric system is not total, since there is a difference which consists in, apart from the flow of momentum-mass and heat there is also convection of magnetic flow and energy. Thus, in fig. 2 we have to point out the following: the successive phases of the magnetotail, is the time evolution of intrinsic instability of the system, while the phases corresponding to the hydrodynamic flow are related to the gradual increase of the velocity of the external flow.*

*Our inability to give an adequate deterministic and local description of the magnetospheric dynamics supports the hypothesis of this study. Namely, that chaos theory could represent a convincing hypothesis concerning the magnetopsheric dynamics as a whole or from a single perspective. Furthermore, a result that enforces this hypothesis is the amplification of turbulence in the phase of the magnerospheric substorm development [28] together with the strengthening of the magnetospheric system with the solar wind, through the reinforcement of the daily magnetic reconnection.*

*In addition, the question which is to be studied through chaos theory is the following: whether the transition of the magnetospheric system from the stable state to the development phase and afterwards to the bursting phase of the magnetospheric storms, corresponds to a sequential transition from simple attractors (stable state – development phase) to chaotic attractors (bursting phase of magnetosheric storms), where small perturbations are amplified.*

### *Conclusions*

*The previous description, concerning the interaction of the solar wind and the magnetospheric field of a planet, emphasizing to Earth's magnetosphere, showed, among others, the current problems of the magnetospheric physics. These problems concern: The sources and especially the supply mechanism of the magnetosphere with solar and ionospheric plasma, the transportation of mass and energy inside the magnetosphere as well as in its environment, during calm periods or magnetospheric storms, the storage of the tranferred plasma and energy and the dissipation mechanism during the magnetospheric storms, the activation and the propagation of energetic*

*particles which cause the magnetospheric bursts, the development of plasma_instabilities (magnetohydrodynamic micro- and macro-instabilities) and its relation to the magnetospheric dynamics in its different phases. Finally, the hypothesis of the strange attractor dynamics was introduced in order to achieve a holistic understanding of the above processes.*

**References**

1. Prigozine I., (1977), Time, structure and fluctuations, Nobel Lecture.; Prigogine, Ilya (1980). From Being To Becoming. Freeman. ISBN 0716711079.

2. Nicolis, G. & Prigogine, I. [1977], Self-organization in nonequilibrium systems: From dissipative structures to order through fluctuations", Willey, New York.; Nicolis G., (1979), Irreversible Thermodynamics, Rep. Prog. Phys., 42, 227-267.; Nicolis, G. [1986] "Dissipative Systems", Rep. Prog. Phys., 49, p. 873.

3. Baker, D.N. et al. (1990), The evolution from weak to strong geomagnetic activity: An interpretation in terms of deterministic chaos, Geophys. Res. Lett. 17, 41.

4. Vassiliadis, D., et al. (1990), Low-dimensional chaos in magnetospheric activity from AE time series, Geophys. Res. Lett. 17, 1841-1844.

5. Pavlos, G. P. et al., Evidence for chaotic dynamics in outer solar plasma and the earth magnetopause, in Chaotic Dynamics: Theory and Practice, edited by A. Bountis, pp. 327–339. Plenum, New York, 1992a.; Pavlos, G. P. et al., Evidence for strange attractor structures in space plasmas, Ann. Geophys. 10, 309–322, 1992b.; Pavlos G.P. et al., Evidence for Strange Attractor Structures in Space Plasmas, Ann. Geophysicae 10, 309-322, EGS-Springer Verlag, 1992c; Pavlos G.P. et al., Chaos and magnetospheric dynamics, Nonlin. Proc. Geophys., 1, 124–135, 1994a.; Pavlos G.P. et al., Nonlinear analysis of Magnetospheric data, Part I. Geometric characteristics of the AE index time series and comparison with nonlinear surrogate data, Nonlin. Proc. Geophys., 6, 51-65, 1999a.; Pavlos G.P. et al., Nonlinear analysis of Magnetospheric data, Part II. Dynamical characteristics of the AE index time series and comparison with nonlinear surrogate data, Nonlin. Proc. Geophys., 6, 79-98, 1999b.; Pavlos G.P. et al., Comments and new results about the magnetospheric chaos hypothesis, Nonlin. Proc. Geophys., 6, 99-127, 1999c.

6. Pavlos G.P. et al., Chaotic Analysis of a time series compased by Seismic Events occurred in Japan, International Journal of Bifurcation and Chaos, 4, 87-98, 1994b.; Pavlos, G.P. et al., Self Organized Criticality or/and Low Dimensional Chaos in Earthquake Processes. Theory and Practice in Hellenic Region, Nonlinear Dynamics in Geosciences, Eds. A.A. Tsonis and J.B. Elsner, pp. 235-259, Springer 2007.; Pavlos, G.P. et al., Self Organized Criticality (SOC) and Chaos Behavior in the Brain Activity, 10th Volume of the series Order and Chaos, Eds. Bountis A. and Pnevmatikos S., University of Patras Press, pp. 145-155, 2008.

7. Iliopoulos A.C., G.P. Pavlos and M.A. Athanasiou, Spatiotemporal Chaos into the Hellenic Seismogenesis: Evidence for a Global Seismic Strange Attractor, Nonlinear Phenomena in Complex Systems, Vol. 11, N. 2, pp. 274-279, 2008.; Iliopoulos A.C. and G.P. Pavlos, "Global Low Dimensional Seismic Chaos in the Hellenic Region", International Journal of Bifurcation and Chaos, 20(7), 2071-2095, 2010.

8. Karakatsanis L.P. and G.P. Pavlos, Self Organized Criticality and Chaos into the Solar Activity, Nonlinear Phenomena in Complex Systems, Vol. 11, N. 2, pp. 280-284, 2008.; Karakatsanis, L.P., Pavlos, G.P., Iliopoulos, A.C., Tsoutsouras, V.G., Evidence for Coexistence, Intermittent Turbulence and Low-Dimensional Chaos Processes in the Solar Flare Dynamics, Modern Challenges in Nonlinear Plasma Physics, American Institute of Physics Conference Proceedings, Vol. 1320, Eds. D. Vassiliadis, S.F. Fung, X. Shao, I.A. Daglis, J.D. Huba, pp. 55-64, Melville, New York 2010.

9. Tsoutsouras, V. Et al., Simulation of healthy and epileptiform brain activity using cellular automata, International Journal of Bifurcation and Chaos, accepted for publication, 2012.

10. Pavlos, G.P. et al.., First and second order non-equilibrium phase transition and evidence for non-extensive Tsallis statistics in Earth's magnetosphere, Physica A, Vol. 390, issue 15, pp. 2819-2839, 1 August 2011.; Pavlos, G.P, Complexity in Theory and Practice: Toward the Unification of Non-equilibrium Physical Processes, Chaotic Modeling and Simulation (CMSIM), issue 1, pp. 123-145, January 2012.; Pavlos, G.P., Understanding the Multi-scale and Multi-fractal Dynamics of Space Plasmas through Tsallis Non-Extensive Statistical Theory, (submitted for publication), Arxiv preprint, arXiv: arXiv:1203.4003, 2012.; Pavlos, G.P. et al., Tsallis statistics and Magnetospheric Self-Organization, Physica A, Vol. 391, issue 11, pp. 3069-3080, June 2012.

11. Karakatsanis, L.P., Pavlos, G.P., Sfiris, D.S., Universality of first and second order phase transition in solar activity. Evidence for non-extensive Tsallis statistics, International Journal of Bifurcation and Chaos, accepted for publication, 2012.

12. Iliopoulos, A.C., Pavlos, G.P., Papadimitriou, E.E., Sfiris, D.S., Chaos, Self Organized Criticality, intermittent turbulence and non-extensivity revealed from seismogenesis in North Aegean area, International Journal of Bifurcation and Chaos, accepted for publication, 2012.

13. Chang, T. (1992), Low-Dimensional Behavior and Symmetry Braking of Stochastic Systems near Criticality – Can these Effects be observed in Space and in the Laboratory," IEEE 20(6), 691-694.; Chang T. (1999), Self-organized criticality, multi-fractal spectra, sporadic localized reconnections and intermittent turbulence in the magnetotail", *Physics of Plasmas* 6(11), 4137-4145.

14. Consolini, G. et al., Multifractal Structure of Auroral Electrojet Index Data, Phys. Rev. Lett. 76, 4082–4085, 1996.; Consolini, G., [1997] "Sandpile cellular automata and magnetospheric dynamics", 8th GIFCO Conference: Cosmic physics in the year 2000, p. 123 - 126.; Consolini G., and Chang T., Magnetic Field Topology and Criticality in Geotail Dynamics: Relevance to Substorm Phenomena, Space Sc. Rev., 95, 309-321, 2001.; Consolini G. et al., On the magnetic field fluctuations during magnetospheric tail current disruption: A statistical approach, J. Geoph. Research, 110, A07202, 2005.

15. Sharma A. S. et al. (2001), Substorm as Nonequilibrium Transitions of the Magnetosphere, J. Atmos. Solar. Terr. Phys., 63, 1399.; Sharma, A.S., Baker, D.N. and Borovsky, J.E., Nonequilibrium phenomena in the magnetosphere, (eds.) Sharma, A.S. and Kaw, P.K., Nonequilibrium Phenomena in Plasmas, 3-22, 2005.; Sharma, A.S., Ukhorski, A.Y. and Sitnov, M.I., Global and multiscale phenomena of the magnetosphere, (eds.) Sharma, A.S. and Kaw, P.K., Nonequilibrium Phenomena in Plasmas, 117-144, 2005.

16. Sitnov M.I. et al. (2001), Modeling substorm dynamics of the magnetosphere: From self-organization and self-organized criticality to nonequilibrium phase transitions, Physical Review E, Vol. 65, 016116.

17. Nottale, L. [1993] "Fractal Space-Time and Micro-physics", Towards a Theory of Scale Relativity, eds. World Scientific.; Nottale, L. [1994] "Scale Relativity, Fractal Space-time and Quantum Mechanics", Chaos, Solitons & Fractals, 4, p. 361.; Nottale, L. [1996] "Scale Relativity and Fractal Space-Time: Applications to Quantum Physics, Cosmology and Chaotic Systems", Chaos, Solitons & Fractals, 7, p. 877.; Nottale L., The Theory of Scale

Relativity: Non-Differentiable Geometry and Fractal Space-Time, Aip Conference Proceedings 2004, 718(20), p.68-95.; Nottale L., (2006), Fractal space – time, non – differentiable and scale relativity, Invited contribution for the Jubilee of Benoit mandelbrot

18. Castro, C. [2005] ", On non-extensive statistics, chaos and fractal strings", Physica A 347, p. 184.

19. Goldfain, E., Fractional dynamics, Cantorian space-time and the gauge hierarchy problem, Chaos, Solitons and Fractals, 22, 513-520, 2004.; Goldfain, E., Complex Dynamics and the High-energy Regime of Quantum Field Theory, Int. J. of Nonlinear Sciences and Numerical Simulation, 6(3), 223-234, 2005.; Goldfain, E., Chaotic Dynamics of the Renormalization Group Flow and Standard Model Parameters, Int. J. of Nonlinear Science, 3, 170-180, 2007.; Goldfain, E., Fractional dynamics and the TeV regime of field theory, Communications in Nonlinear Science and Numerical Simulation, 13, 666-676, 2008.

20. El Naschie, M.S. [1998] "Superstrings, Knots, and Noncommutative Geometry in E(infinity) Space", Int. Journal of Theoretical Physics, 37, p. 2935.; El Naschie, M.S. [2004] "A review of E infinity theory and the mass spectrum of high energy particle physics", Chaos, Solitons & Fractals, 19, p. 209-236.; El Naschie, M.S. [2005] "Einstein's dream and fractal geometry", Chaos, Solitons & Fractals, 24, p. 1.; El Naschie, M.S. [2006] "Elementary prerequisites for E-infinity (Recommended background readings in nonlinear dynamics, geometry and topology)", Chaos, Solitons & Fractals, 30, p. 579.

21. Cresson, J., Non-differentiable variational principles, J. Math. Anal. Appl., 307, 48-64, 2005.; Cresson, J. and Greff, I., Non-differentiable embedding of Lagrangian systems and partial differential equations, J. Math. Anal. Appl., 384, 626-646, 2011.

22. Zelenyi, L. M. & Milovanov, A. V., Fractal Properties of Sunspots, Soviet Astronomy Letters, 17, p. 425, 1991.; b) Zelenyi, L. M. & Milovanov, A. V., Fractal topology and strange kinetics: from percolation theory to problems in cosmic electrodynamics, Pysics-Uspekhi, 47(8), 749-788, 2004.; Zelenyi, L.M., Malova, H.V., Artemyev, A.V., Popov, V.Yu. and Petrukovich, A.A., Thin Current Sheets in Collisionless Plasma: Equilibrium structure, Plasma Instabilities, and Particle Acceleration, Plasma Physics Reports, 37(2), 118-160, 2011.

23. Milovanov, A.V. and Zelenyi, L.M., Functional background of the Tsallis entropy: "coarse-grained" systems and "kappa" distribution functions, Nonlinear Processes in Geophysics, 7, 211-221, 2000.; Milovanov, A.V. and Rasmusse, J.J., Fractional generalization of the Ginzburg-Landaou equation: an unconventional approach to critical phenomena in complex media, Physics Letters A377, 75-80, 2005.; Milovanov, A.V., Dynamic Polarization random walk model and fishbone-like instability for self-organized critical systems, New Journal of Physics, 13, 043034, 2011.

24. *Eckmann, J.P. and D. Ruelle*, Rev. Mod. Phys. 57, 617 (*1985*)

25. *Buchner*, J. and L.M. *Zelenyi*, J. Geophys. Res., 92, 13456, *1987.*

26. Prigogine, I. (1980). *From Being To Becoming*. Freeman. ISBN 0716711079.

27. Landau, L.D. and E.M. Lifshitz (1987). *Fluid Mechanics.* **Vol. 6** (2nd ed.). Butterworth-Heinemann. ISBN 978-0-080-33933-7.

28. Huba, J.D. et al., Geophys. Res. Lett., 4, 125, 1977.